\documentclass{PoS}

\title{Flavour constraints on beyond the Standard Model scenarios}

\ShortTitle{Flavour constraints on beyond the Standard Model scenarios}

\author{\speaker{Farvah Mahmoudi}\\ 
	  Clermont Universit\'e, Universit\'e Blaise Pascal, CNRS/IN2P3, LPC, BP 10448, F--63000 Clermont-Ferrand, France\\ 
CERN Theory Division, Physics Department, CH-1211 Geneva 23, Switzerland \\ 
E-mail: \email{mahmoudi@in2p3.fr}}

\abstract{The interplay of flavour and collider physics is entering a new era with the start-up of the LHC. During the past few years rare $B$ decays and in particular $b \to s \gamma$ transitions have been extensively used and provided exciting opportunities for mapping possible routes beyond the SM. Flavour constraints play in this manner a complementary role to the direct searches.
Here we present an overview of the existing flavour constraints on various models and show examples of comparison with the LHC discovery potentials. The SuperIso program which is dedicated to flavour physics observable calculations is also described briefly.}

\FullConference{35th International Conference of High Energy Physics\\
         July 22-28, 2010\\
         Paris, France}

\begin{document}

\section{Introduction}

In addition to direct searches for new physics and new effects, indirect searches play an important and complementary role in the quest for physics beyond the Standard Model (SM). The most commonly used indirect constraints originate from flavour physics observables, cosmological data and relic density, electroweak precision tests and anomalous magnetic moment of the muon. 
Precise experimental measurements and theoretical predictions have been achieved for the $B$ meson systems in the past decade \cite{Hurth:2010tk} and stringent constraints due to sizeable new physics contributions to many observables \cite{Carena:2006ai,eriksson,Mahmoudi:2009zx} can be obtained. 
Here we investigate flavour data constraints on two new physics scenarios, namely the two Higgs doublet extension (THDM) and the supersymmetric extension (SUSY) of the SM.

\section{Flavour observables}

Flavour observables can be classified in different categories, such as radiative penguin decays, electroweak penguin decays, neutrino modes and meson mixings.

The inclusive branching ratio of $B \to X_s \gamma$ and the isospin asymmetry of $B \to K^* \gamma$ are the most important observables in the first category. Since $b \to s \gamma$ transition occurs first at one-loop level in the SM, new physics contributions can be of comparable
magnitude. Here penguin loops involve the $W$ boson in the Standard Model, and in addition charged Higgs boson, chargino, neutralino and gluino in the MSSM. The charged Higgs loop always adds constructively to the SM penguin. Thus, BR($B \to X_s \gamma$) is an effective tool to probe the THDM scenario. Chargino loops however can add constructively or destructively. If the interference is positive, it results in a great enhancement in the BR($B \to X_s \gamma$), which becomes therefore a powerful observable. On the other hand, if the interference is negative, the other interesting observable which opens up is the degree of isospin asymmetry in the exclusive decay of $B \to K^* \gamma$.

The most relevant observables in electroweak penguin decays are the branching ratio of $B_s \to \mu^+ \mu^-$, branching ratios and forward-backward asymmetries in $B \to X_s \ell^+ \ell^-$ and $B \to K^{(*)} \mu^+ \mu^-$ decays. In SUSY in the large $\tan\beta$ regime, these rare decays are dominated by the exchange of neutral Higgs bosons and substantial enhancements in the branching ratios can be expected. 

Finally in the neutrino mode category, branching ratios of $B_u \to \tau \nu_\tau$, $B \to D \tau \nu_\tau$, $D_s \to \tau \nu_\tau$, $D_s \to \mu \nu_\mu$, $K \to \mu \nu_\mu$, as well as double ratios of leptonic decays are the most important observables.
These decays can be mediated by a charged Higgs boson already at tree level in annihilation processes and therefore are very sensitive to the charged Higgs sector.

\begin{figure}[!t]
\begin{center}
\vspace*{-0.2cm}\includegraphics[width=5.cm]{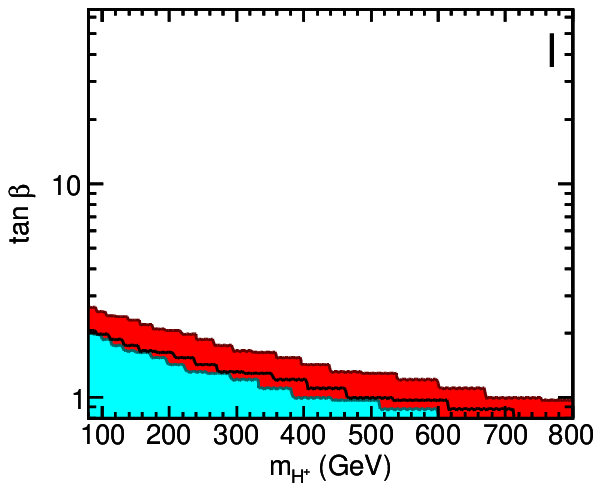}\includegraphics[width=5.cm]{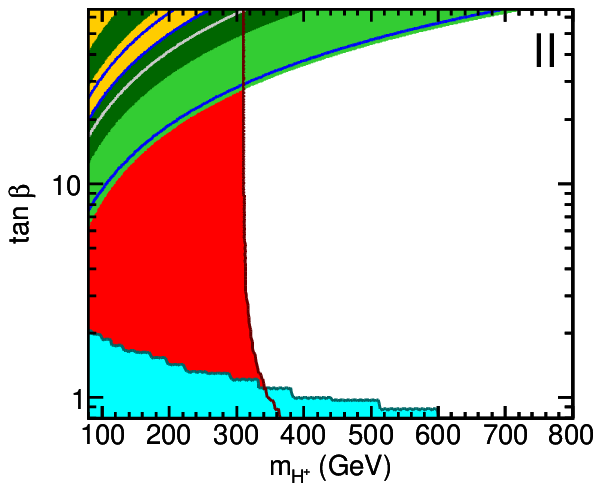}\\
\includegraphics[width=5.cm]{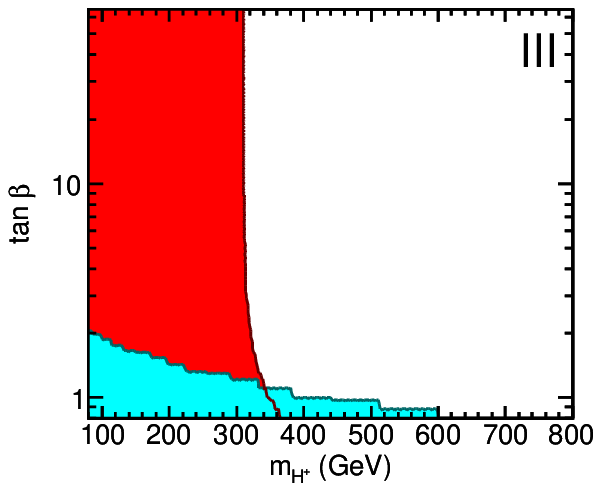}\includegraphics[width=5.cm]{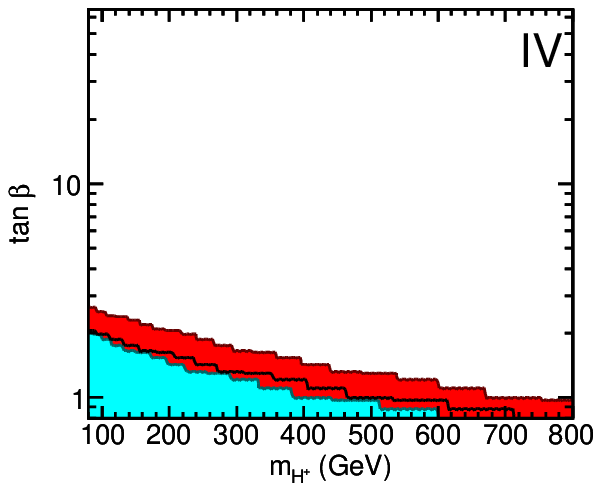}
\end{center}
\caption{Excluded regions of the ($m_{H^+},\tan\beta$) parameter space for $Z_2$-symmetric THDM types. The colour coding is as follows: $B\to X_s\gamma$ (red), $\Delta_{0-}$ (black contour), $\Delta M_{B_d}$ (cyan), $B_u\to\tau\nu_\tau$ (blue), $B\to D\tau\nu_\tau$ (yellow), $K\to\mu\nu_\mu$ (grey contour), $D_s\to\tau\nu_\tau$ (light green), and $D_s\to\mu\nu_\mu$ (dark green).\label{2hdm}}
\end{figure}

Figure \ref{2hdm} presents the combined constraints on the four $Z_2$-symmetric THDM types \cite{Mahmoudi:2009zx}. We first note the exclusion of low $\tan\beta < 1$ in all four models for $m_{H^+} < 500$ GeV as a result of three observables: BR$(B \to X_s \gamma)$, isospin asymmetry, and $\Delta M_{B_d}$. The constraints at low $\tan\beta$ are similar between the models, since the couplings to the up-type quarks are universal. In THDM types II and III, which share the same coupling pattern for the quarks, there exists a $\tan\beta$-independent lower limit of $m_{H^+} \ge 300$ GeV imposed by BR$(B \to X_s \gamma)$. No generic lower limit on $m_{H^+}$ is found in type I and type IV models. Constraints for high $\tan\beta$ are only obtained in the type II model since the leptonic and semi-leptonic observables require $\tan\beta$-enhanced couplings for the contributions to be interesting.

\begin{figure}[!t]
\begin{center}
\includegraphics[width=6.5cm]{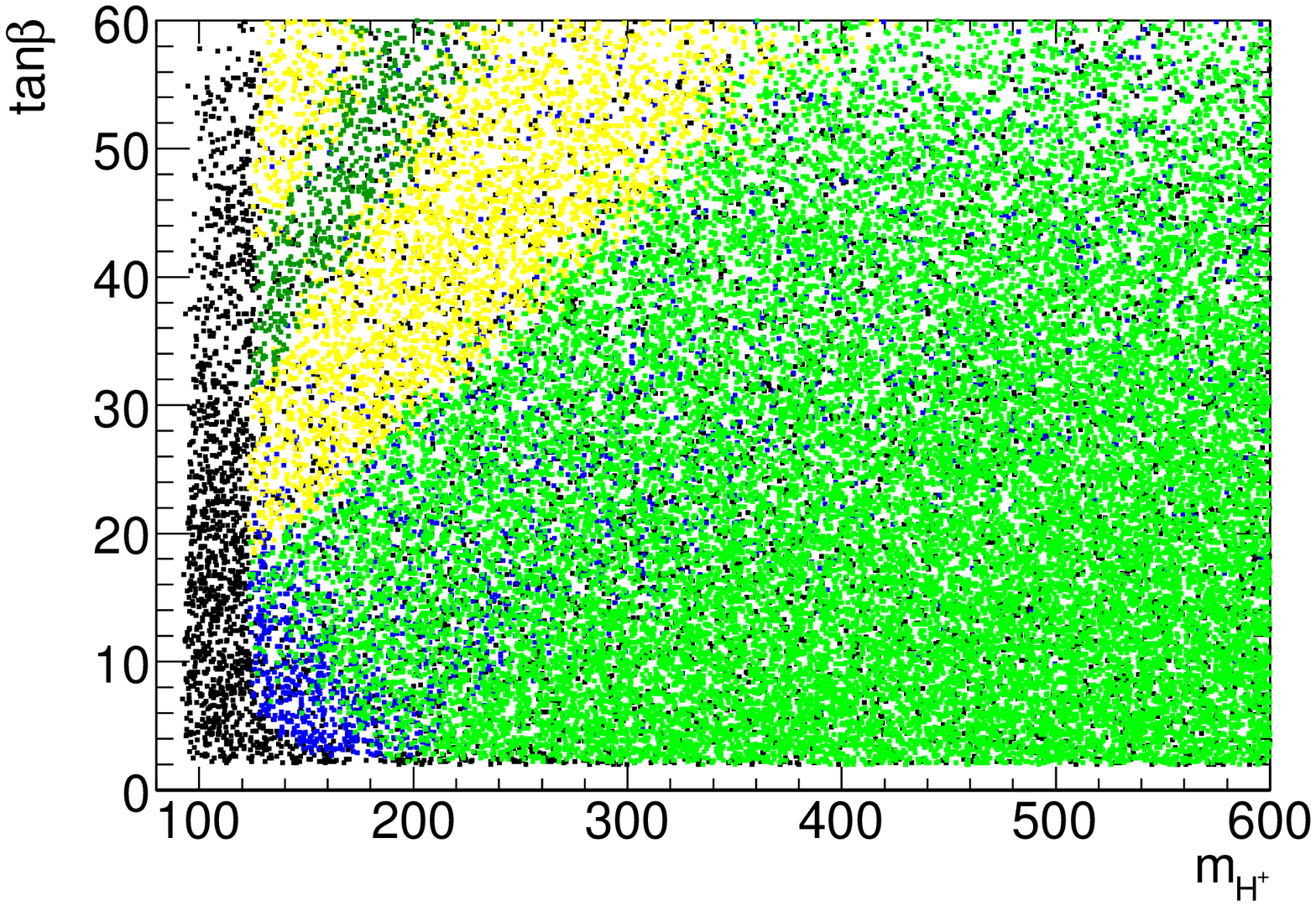}\includegraphics[width=6.5cm]{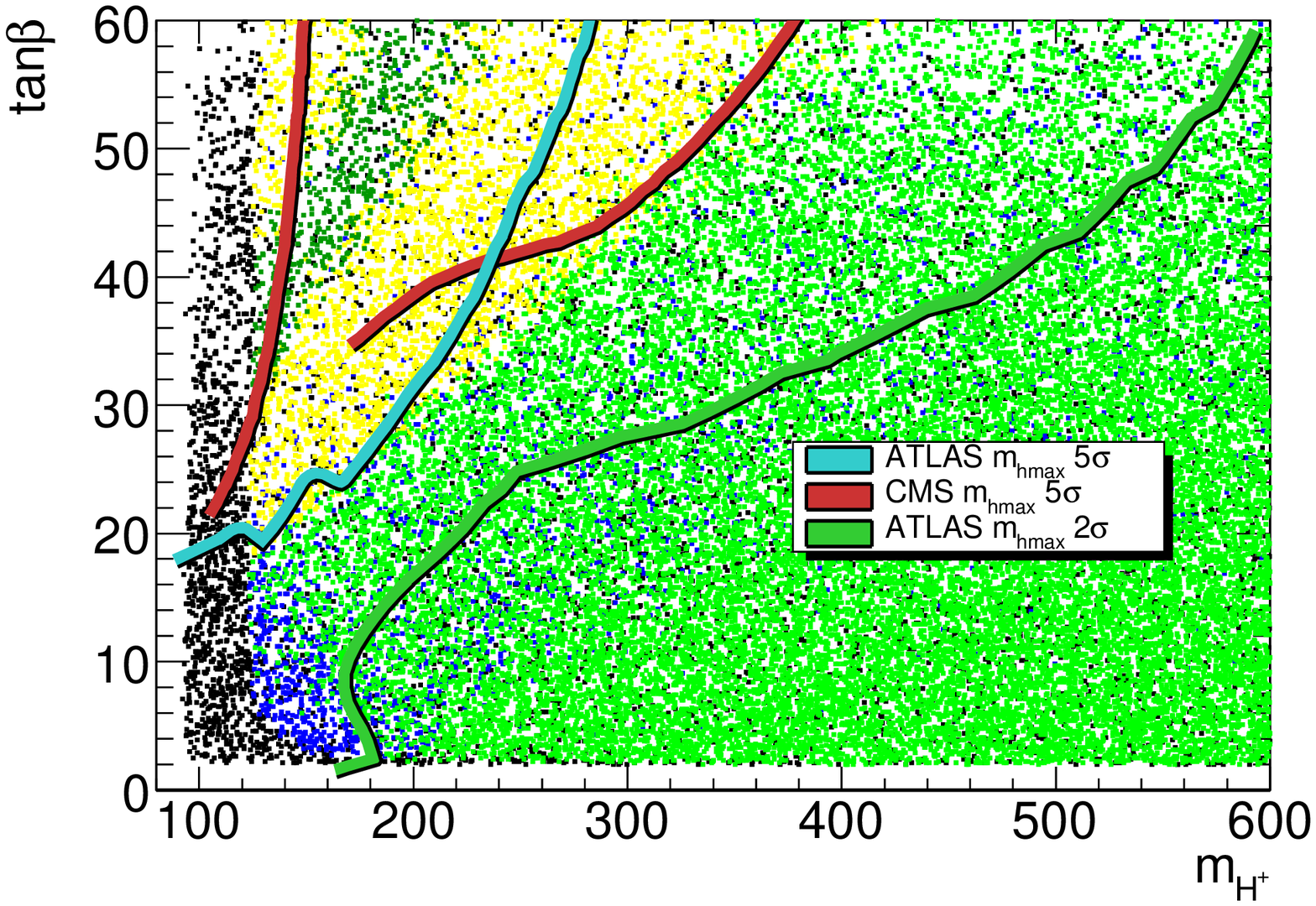}\includegraphics[clip,bb=0 25 48 141,width=1.8cm]{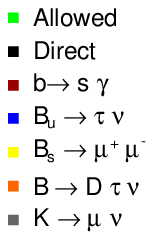}
\end{center}
\caption{Combined exclusion in NUHM models by different constraints. The constraints are applied in the order they appear in the legend, and the colour coding corresponds to the first constraint by which a point is excluded. All points have $\mu > 0$ and a neutral LSP. In the right plot the experimental discovery contours for ATLAS and CMS are superimposed.\label{nuhm}}
\end{figure}

Figure \ref{nuhm} shows a combination of constraints applied to the NUHM parameter space \cite{eriksson}. Here we see that charged Higgs masses down to $m_{H^+}\simeq 135$~GeV can be accommodated, with the lowest masses allowed for intermediate $\tan\beta\sim 7$--$15$. For higher $\tan\beta$, the combined constraints follow the exclusion by $B_u\to\tau\nu_\tau$. A comparison between the experimental results and the NUHM points is also presented. This comparison illustrates that most of the indirect constraints are relevant in the same parameter space regions where the charged Higgs production cross section at the LHC is the largest. The discovery of the charged Higgs would therefore serve as an indication of a non-minimal model being
realized in nature.

\begin{figure}[!t]  
\begin{center}
\includegraphics[width=6cm]{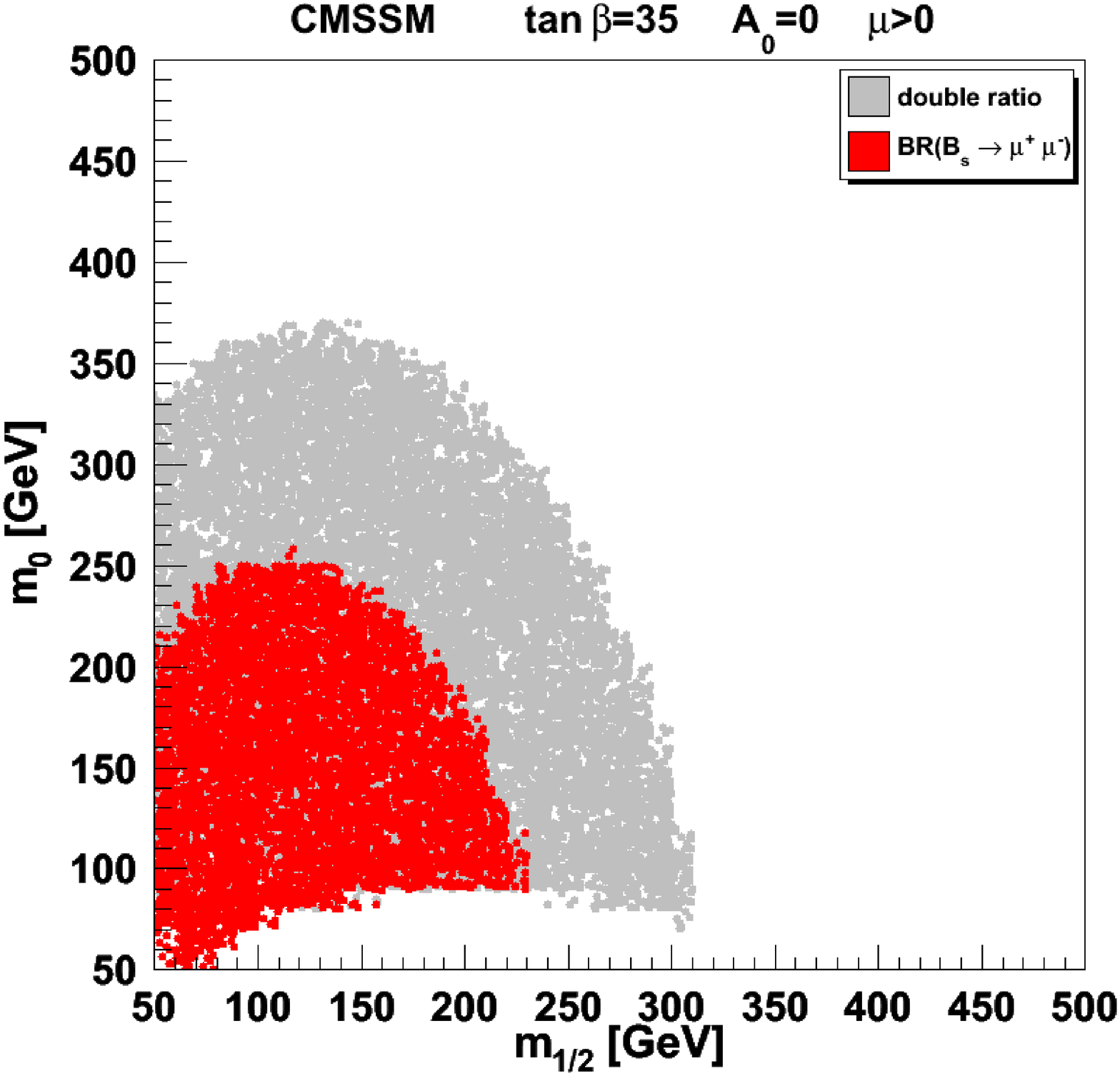}\includegraphics[width=6cm]{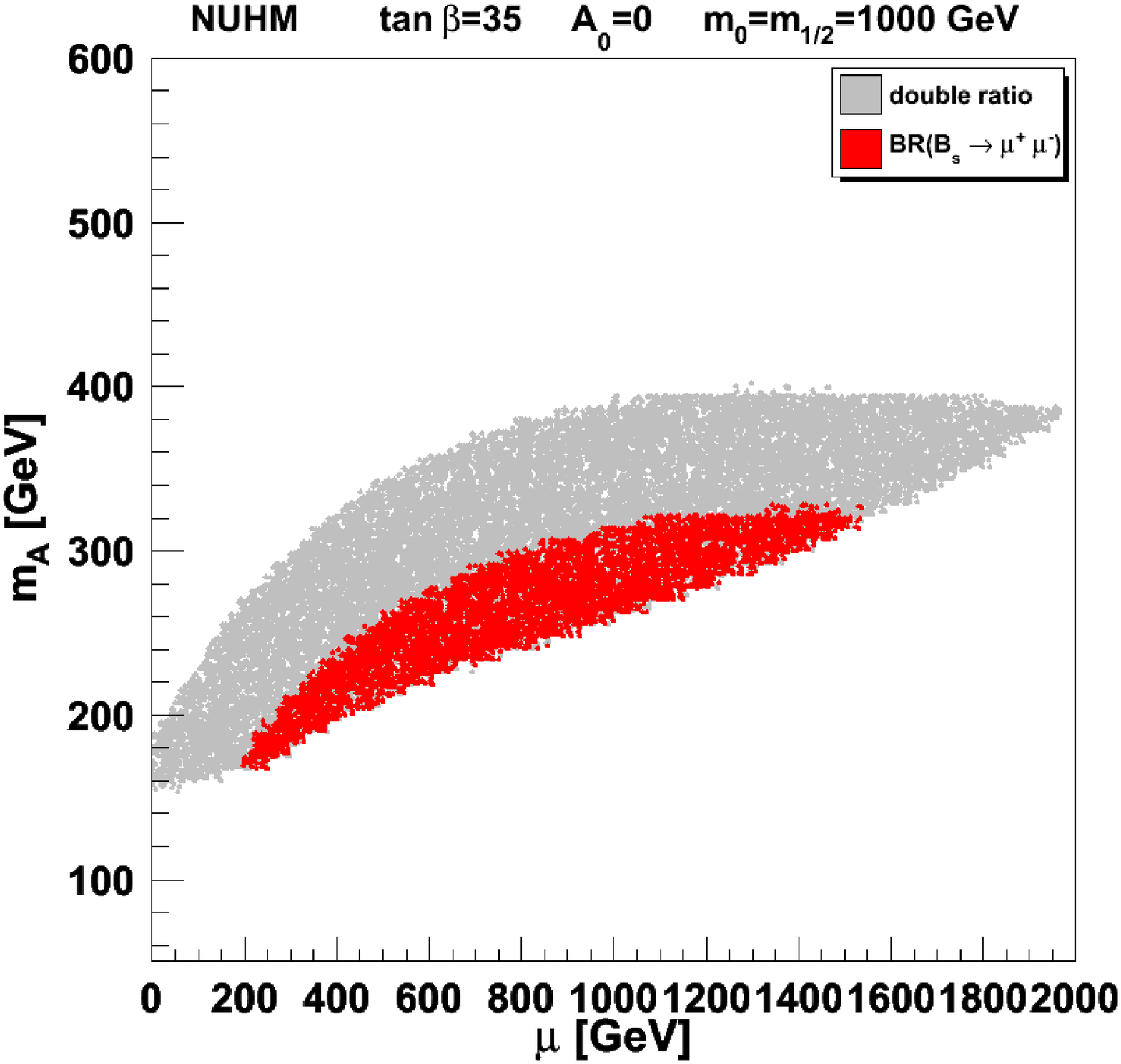}
\end{center}
\caption{In the left panel, CMSSM parameter plane $(m_{1/2},m_0)$ with $\tan\beta = 35$, $A_0=0$ and $\mu >0$. In the right panel, NUHM parameter plane $(\mu,m_A)$ with $\tan\beta = 35$, $A_0=0$ and $m_0=m_{1/2}=1000$ GeV. The points in red are excluded at $95\%$ C.L. by BR$(B_s\to\mu^+\mu^-)$ and in grey by the double ratio (Eq. (2.1)).\label{dr}}
\end{figure}

Most of the leptonic observables are subject to uncertainties from decay constants. In order to remove such uncertainties it is possible to define double ratios of leptonic decays in a way to cancel the dependency on the decay constants \cite{doubleratio}. In figure \ref{dr} we compare the constraints obtained by BR$(B_s\to\mu^+\mu^-)$ and by the double ratio
\begin{equation}
\left(\frac{\mathrm{BR}(B_s\to\mu^+\mu^-)}{\mathrm{BR}(B_u \to \tau \nu_\tau)}\right) \Big/ \left(\frac{\mathrm{BR}(D_s\to\tau\nu_\tau)}{\mathrm{BR}(D\to\mu\nu_\mu)}\right) 
\label{eqdr}
\end{equation}
in the CMSSM and in the NUHM scenarios. 
It can be seen that the double ratio, being a combination of four different constraints, can be
more constraining than the branching ratio of $B_s\to\mu^+\mu^-$ taken individually. 
Importantly, contrary to $B_s\to\mu^+\mu^-$, the double ratio does not depend on lattice inputs, but instead depends on $|V_{ub}|$, whose magnitude is already constrained from various distinct experimental methods. Hence the double ratio is an important alternative to BR$(B_s\to\mu^+\mu^-)$ as a probe of the SUSY parameter space.

\section{SuperIso program}

SuperIso \cite{superiso} is a public C program dedicated mostly to the calculation of flavour physics observables. 
The calculations are done in various models, such as SM, THDM, MSSM and NMSSM with minimal flavour violation.
A broad set of flavour physics observables is implemented in SuperIso. This includes the branching ratio of $B \to X_s \gamma$, isospin asymmetry of $B \to K^* \gamma$, branching ratio of $B_s \to \mu^+ \mu^-$, branching ratios of $B_s \to X_s \mu^+ \mu^-$, $B_s \to K^* \mu^+ \mu^-$, $B_s \to K \mu^+ \mu^-$ and the forward backward asymmetries in these decays, branching ratio of $B_u \to \tau \nu_\tau$, branching ratio of $B \to D \tau \nu_\tau$, branching ratio of $K \to \mu \nu_\mu$, branching ratio of $D \to \mu \nu_\mu$, and the branching ratios of $D_s \to \tau \nu_\tau$ and $D_s \to \mu \nu_\mu$.
The calculation of the anomalous magnetic moment of the muon is also implemented in the program. SuperIso uses a SUSY Les Houches Accord (SLHA) file \cite{slha} as input, which can be either generated automatically by the program via a call to a spectrum generator or provided by the user.
The program is able to perform the calculations automatically for different SUSY breaking scenarios. 
An extension of SuperIso including the relic density calculation, SuperIso Relic, is also available publicly \cite{superiso_relic}. 
Finally, in SuperIso we make use of the Flavour Les Houches Accord (FLHA) \cite{Mahmoudi:2010iz}, the newly developed standard for flavour related quantities, and the program provides an FLHA output file as well.

\section{Conclusion}

Indirect constraints and in particular those from flavour physics are essential
to restrict new physics parameters as we have seen here. The information obtained from these low energy observables combined with LHC data will open the door to a very rich phenomenology and would help us step forward toward a deeper understanding of the governing physics. Here we showed a few examples of possible analyses but the same methods can of course be generalized to more new physics
scenarios.


\end{document}